\documentclass[12pt]{article}
\usepackage{amssymb}
\usepackage{amsmath}
\usepackage{hyperref}
\begin{document}
\title{\bf Einstein-Yang-Mills AdS Black Brane Solution in Massive Gravity and Viscosity Bound }
\author{Mehdi Sadeghi\thanks{ mehdi.sadeghi@abru.ac.ir}  \hspace{2mm} \\
       {\small {\em Department of Sciences, University of Ayatollah Ozma Borujerdi, Borujerd, Lorestan, Iran}}\\
       }
\date{\today}
\maketitle

\abstract{We introduce the Einstein-Yang-Mills AdS black brane solution in context of massive gravity. The ratio of shear viscosity to entropy density is calculated for this solution. This value violates the KSS bound if we apply the Dirichlet boundary and regularity on the horizon conditions.}\\



\section{Introduction} \label{intro}

\indent   General theory of relativity introduced by Albert Einstein in 1915 is a theory that graviton is massless within it. This theory predicts the gravitational waves which observed by advanced LIGO in 2016, but there are some phenomena that GR cannot explain them  including the current acceleration of the universe, the cosmological constant problem, dark energy and dark matter. In recent decades, GR had been generalized to explain these problems such as massive
gravity\cite{deRham:2010kj}, bimetric gravity\cite{Hassan:2011ea}, scalar-tensor gravity\cite{Ref03} and modified gravity \cite{Ref04}-\cite{Ref06}. On the other hand, the hierarchy problem and brane-world gravity solutions suggest that graviton isn't massless.\\

 Massive gravity helps us to study the quantum gravity effects and this theory includes some interesting properties: (i) it could explain the accelerated expansion of the universe without considering the dark energy, (ii) the graviton behaves like a lattice excitation and exhibits a Drude peak in this theory, (iii) current experimental data from the observation of gravitational waves by advanced LIGO requires the graviton mass\cite{Abbott:2016blz}.\\

Massive gravity introduced by Fierz-Pauli \cite{Fierz:1939ix} suffers from vDVZ (van Dam-Veltman-Zakharov) discontinuity problem. To resolve this problem, it must be considered in a nonlinear framework according to Vainshtein proposal. This proposal contains Boulware-Deser ghost. Finally de Rham, Gabadadze and Tolley (dRGT) solve this problem\cite{deRham:2010kj}.\\

  Gauge/Gravity duality  \cite{Ref1}-\cite{Ref4} relates two different types of theories: gravity in $(d+1)$-dimension and gauge theory in $d$-dimension. Perturbation theory is not applicable to the strongly coupled gauge theories but gauge/gravity  duality opens a window for solving these theories by introducing a dictionary which can translate the information of strongly coupled gauge theory into a weakly gravity theory and vice versa. Gauge theory lives on the boundary of $AdS_5$ and gravity on the bulk of $AdS_5$ in this duality, thus this forced us the background to be Anti-de sitter spacetime. In the long wavelength regime this duality leads to fluid/gravity duality \cite{Ref5}-\cite{Ref10}. Any fluid is characterized by some transport coefficients. One of these transport  coefficients is the shear viscosity. This duality  is a powerful method  for the the computation of transport coefficients in strongly coupled gauge theories in the hydrodynamic limit. There is three ways to calculate this coefficient: pole method, Green-Kubo formula and membrane paradigm. In section \ref{sec4} of this paper, shear viscosity is calculated via Green-Kubo formula.\\
 In the Green-Kubo formula approach, transport coefficients of plasma are related to their thermal correlators and we use gauge/gravity duality for finding this correlator \cite{Ref13}-\cite{Ref19}.
 \begin{equation}
\eta =\mathop{\lim }\limits_{\omega \to 0} \frac{1}{2\omega } \int dt\,  d\vec{x}\, e^{i\omega t} \left\langle [T_{y}^{x} (x),T_{y}^{x} (0)]\right\rangle =-\mathop{\lim }\limits_{\omega \, \to \, 0} \frac{1}{\omega } \Im G_{y\, \, y}^{x\, \, x} (\omega ,\vec{0}).
\end{equation}
The ratio of shear viscosity to entropy density is proportional to the inverse square coupling  of quantum thermal gauge theory. It means the stronger the coupling, the weaker the shear viscosity per entropy density.\\
The study of Quark Gluon Plasma (QGP) arises from the fact that after Big Bang the universe was filled with very hot and dense soup, known as QGP, which is strongly coupled. In laboratory, it is created by head-on collision between heavy ions such as gold or lead nuclei. The lower  bound of  $\frac{\eta}{s}$  is related to QGP in all fluids in the nature due to its strongly coupled charactrestic. The KSS bound supports both string theory outcomes \cite{Ref17}  and quark-gluon plasma experimental data \cite{Shen:2015msa}. \\\\
In this paper we consider massive gravity in the presence of Yang-Mills gauge field and introduce the black-brane solution. This model can be considered as a generalization of the Einstein-Hilbert model for studying the unknown part of the universe, dark matter and dark energy. Finally, we check the Dirichlet boundary and regularity on the horizon conditions for the value of $\frac{\eta}{s}$.
\section{Einstein-Yang-Mills AdS Black Brane Solution in Massive Gravity}
\label{sec2}

\indent The action of 5-dimensional Einstein-massive gravity with negative cosmological constant in the presence of Yang-Mills source is as below,

\begin{equation}\label{Action}
S=\int d^{5}  x\sqrt{-g} \Big(R-2\Lambda-\gamma _{ab} F_{\mu \nu }^{(a)} F^{(b)\, \, \mu \nu }+m^2\sum_{i=1}^4{c_{i}\mathcal{U}_i(g,f)}\Big),
\end{equation}
where $R$ is the Ricci scalar,$\Lambda=\frac{-6}{\l^2}$ the cosmological constant, $l$ the AdS radius and $F_{\mu \nu }^{(a)} $ the $SO(5,1) $ Yang-Mills gauge field tensor \cite{Sadeghi:2014uqf},
\begin{align} \label{eq03}
F_{\mu \nu }^{(a)} =\partial _{\mu } A_{\nu }^{(a)} -\partial _{\nu } A_{\mu }^{(a)} +\frac{1}{2e} C_{bc}^{a} A_{\mu }^{(b)} A_{\nu }^{(c)} \;\;\; a=1,\, 2,...,N,
\end{align}
where $e$ is the gauge coupling constant, $C$'s are the gauge group structure constants, $A_{\nu }^{(a)} $'s the gauge potentials and $\gamma _{ab} =-\frac{\Gamma _{ab} }{|\det \Gamma _{ab}|^{^{\frac{1}{N} } } }$ the metric tensor of the gauge group in which $\Gamma _{ab} =C_{ad}^{c} C_{bc}^{d}$ and  $|\det \Gamma _{ab}|>0$.\\

In (\ref{Action}), $ c_i $'s are constants and $ \mathcal{U}_i $ are symmetric polynomials of the eigenvalues of the $ 5\times5 $ matrix $ \mathcal{K}^{\mu}_{\nu}=\sqrt{g^{\mu \alpha}f_{\alpha \nu}} $
\begin{align}\label{7}
  & \mathcal{U}_1=[\mathcal{K}]\nonumber\\
  & \mathcal{U}_2=[\mathcal{K}]^2-[\mathcal{K}^2]\nonumber\\
  &\mathcal{U}_3=[\mathcal{K}]^3-3[\mathcal{K}][\mathcal{K}^2]+2[\mathcal{K}^3]\nonumber\\
  & \mathcal{U}_4=[\mathcal{K}]^4-6[\mathcal{K}^2][\mathcal{K}]^2+8[\mathcal{K}^3][\mathcal{K}]+3[\mathcal{K}^2]^2-6[\mathcal{K}^4]
   \end{align}
Variation of the action (\ref{Action}) with respect to the metric tensor $g_{\mu \nu } $ and the Yang-Mills field tensor $F_{\mu\nu }^{(a)} $ leads to,
\begin{align}
&G_{\mu \nu } +\Lambda g_{\mu \nu } =8\pi T_{\mu \nu },\\&
F_{;\nu }^{(a)\mu \nu } =j^{(a)\mu },
\end{align}
where $G_{\mu \nu}$ is the Einstein tensor and  the gauge current and the stress-energy tensor carried by the gauge fields are as follows,
\begin{align}
&T_{\mu \nu } =\frac{1}{4\pi } \gamma _{ab} (F_{\mu }^{(a)\lambda } F_{\nu \lambda }^{(b)} -\frac{1}{4} F^{(a)\lambda \sigma } F_{\lambda \sigma }^{(b)} g_{\mu \nu } )+m^2 \mathcal{\chi}_{\mu \nu},\\&
j^{(a)\nu } \, =\frac{1}{e} C_{bc}^{a} A_{\mu }^{(b)} F^{(c)\mu \nu },\\&
 \mathcal{\chi}_{\mu \nu} =-\frac{c_1}{2}\bigg(\mathcal{U}_1g_{\mu \nu }-\mathcal{K}_{\mu \nu}\bigg)-\frac{c_2}{2}\bigg(\mathcal{U}_2g_{\mu \nu }-2\mathcal{U}_1\mathcal{K}_{\mu \nu}+2\mathcal{K}^2_{\mu \nu}\bigg)-\frac{c_3}{2}\bigg(\mathcal{U}_3g_{\mu \nu }-3\mathcal{U}_2\mathcal{K}_{\mu \nu}\nonumber\\&+6\mathcal{U}_1\mathcal{K}^2_{\mu \nu}-6\mathcal{K}^3_{\mu \nu}\bigg)-\frac{c_4}{2}\bigg(\mathcal{U}_4g_{\mu \nu }-4\mathcal{U}_3\mathcal{K}_{\mu \nu}+12\mathcal{U}_2\mathcal{K}^2_{\mu \nu}-24\mathcal{U}_1\mathcal{K}^3_{\mu \nu}+24\mathcal{K}^4_{\mu \nu}\bigg).\end{align}

$ \mathcal{\chi}_{\mu \nu}$  is the massive term.

\noindent The invariant scalar ${\mathcal F}\equiv \gamma _{ab} F^{(a)\lambda \sigma } F_{\lambda \sigma }^{(b)}$ for the YM fields is,
\begin{equation}
 {\mathcal F}_{YM} =\frac{6e^{2} }{r^{4} } .
\end{equation}
We consider the following metric ansatz for a five-dimensional planar AdS black brane,
\begin{equation}
ds^{2} =-\frac{r^2N(r)^2}{l^2}f(r)dt^{2} +\frac{l^2dr^{2}}{r^2f(r)} +r^2h_{ij}dx^idx^j,
\end{equation}
A generalized version of $ f_{\mu \nu} $ was proposed in {\cite{Cai:2014znn}} with the  form
$ f_{\mu \nu} = diag(0,0,c_0^2h_{ij})$, where $ h_{ij}=\frac{1}{l^2}\delta_{ij} $.

The values of $ \mathcal{U}_i $ are calculated as below,
\begin{align}\label{9}
  & \mathcal{U}_1=\frac{3c_{0}}{r}, \,\,\,  \,\,\, \mathcal{U}_2=\frac{6c_0^2}{r^2},\,\,\,\,\mathcal{U}_3=\frac{6c_0^3}{r^3},\,\,\,\,\mathcal{U}_4=0\nonumber
   \end{align}
Inserting this  ansatz into the action  Eq.(\ref{Action}) yields,
\begin{equation}\label{EOMI}
 I =\int{d^5x\frac{3N(r)}{l^5}\frac{d}{dr}\Bigg[r^4\bigg(1-f(r)-\frac{2e^{2} l^2}{r^4}\ln r\bigg)+ m^2l^2c_0\bigg(\frac{c_1 r^3}{3}+c_0c_2r^2+2c_0^2c_3r\bigg)\Bigg]}
\end{equation}
We can find the equation of motion by variation of $N(r)$ {\cite{Ref22}},
\begin{equation}\nonumber
\frac{d}{dr}\Bigg[r^4\bigg(1-f(r)-\frac{2e^{2} l^2}{r^4}\ln r\bigg)+ m^2l^2\bigg(\frac{c_0c_1 r^3}{3}+c_0^2c_2r^2+2c_0^3c_3r\bigg)\Bigg]=0
\end{equation}
$ f(r)$ is found by solving the following equation,
\begin{equation}\label{f}
r^4\bigg(1-f(r)-\frac{2e^{2} l^2}{r^4}\ln r \bigg)+ m^2l^2\bigg(\frac{c_0c_1 r^3}{3}+c_0^2c_2r^2+2c_0^3c_3r\bigg)=b^4
\end{equation}
 where $r_0$ and $ b$  are  integration constants. $f(r)$ is found by solving Eq.(\ref{f}) which yields,

\begin{equation}
 f(r)=1-\frac{b^4}{r^4}-\frac{2e^{2} l^2}{r^4}\ln r+m^2l^2 \bigg(\frac{c_0c_1}{3r}+\frac{c_0^2c_2}{r^2}+\frac{2c_0^3c_3}{r^3}\bigg).
\end{equation}
 Event horizon is where $f(r_0)=0$ and we can find $b^4$  by applying this condition,
  \begin{equation}
b^4= r_0^4\Bigg[1-\frac{2e^{2}l^2}{r_0^4}\ln r_0+m^2l^2\bigg(\frac{c_0c_1}{3r_0}+\frac{c_0^2c_2}{r_0^2}+\frac{2c_0^3c_3}{r_0^3}\bigg)\Bigg]
 \equiv r_0^4\bigg(1-\frac{2e^{2}l^2}{r_0^4}\ln r_0+\Delta\bigg)\\
\end{equation}
where $\Delta$ is,
 \begin{equation}
\Delta \equiv m^2l^2\bigg(\frac{c_0c_1}{3r_0}+\frac{c_0^2c_2}{r_0^2}+\frac{2c_0^3c_3}{r_0^3}\bigg).
\end{equation}
By substituting $b^4$ in $f(r)$ we will have,
\begin{equation}
f(r)=\frac{l^2}{3r^4}\Big(6 c_0^3 c_3 m^2(r-r_0) +3 c_0^2 c_2  m^2(r^2-r_0^2) + c_0 c_1 m^2(r^3-r_0^3) +\frac{3}{l^2}(r^4-r_0^4)-6e^{2}\ln (\frac{r}{r_0})\Big).
\end{equation}
It is easy to show that  $N(r)$  is constant by  variation of  $f(r)$ . By applying the speed of light  $c=1$ in the boundary, $N$ is found as the following,
\begin{equation}
\lim_{r \to r_{b}}{N^2f(r)}=1\,\,\,\, \to N^2=\frac{1}{1-\frac{2e^2l^2}{r_b^4}\ln (\frac{r_b}{r_0})}\Bigg|_{r_b=\infty}=1
\end{equation}
\noindent Where $r_{b}$ is boundary of AdS.\\
Einstein-Yang Mills solutions are different from Einstein-Maxwell equation in  $d>4$.\\
 Hawking temperature is defined by,
\begin{equation} \label{eq18}
T=\frac{\kappa (r_{0} )}{2\pi } = \frac{1}{2\pi \sqrt{g_{rr} } } \frac{d}{dr} \sqrt{g_{tt} }|_{r=r_{0} }=\frac{1}{4\pi \sqrt{g_{rr} g_{tt} } } \partial _{r} g_{tt}|_{r=r_{0} }
\end{equation}
where, $\kappa (r_{0} )$ is the surface gravity on the event horizon.
In our case temperature is,
\begin{equation} \label{eq19}
T=\frac{r_0}{4\pi l^2}\bigg[4-\frac{2e^2l^2}{ r_0^4}-m^2l^2\bigg(\frac{c_0 c_1}{3 r_0}+\frac{2 c_0^2 c_2}{r_0^2}+\frac{6c_0^3 c_3}{r_0^3}\bigg)\bigg].
\end{equation}
The entropy can be found by using Hawking-Bekenstein formula,
\begin{eqnarray}
A&=&\int d^{3} x \sqrt{-g} |_{r=r_0,t=cte}= \frac{r_{0}^{3} V_{3}}{l^{3}} \nonumber\\
S&=&\frac{A}{4G} =\frac{r_{0}^{3} V_{3} }{4l^{3}G} \nonumber\\
s&=&\frac{S}{V_{3} } =\frac{4\pi r_{0}^{3} }{l^{3}}
\end{eqnarray}
where $V_{3}$ is the volume of the constant $t$ and $r$ hyper-surface with radius $r_{0}$ and in the last line we used $\frac{1}{16\pi G} =1$ so $\frac{1}{4G} =4\pi$.\\

\section{Shear Viscosity to Entropy Density}
 \label{sec4}
The black brane solution is,
 \begin{align}
   & ds^{2} =-\frac{f_1(r)}{l^2}dt^{2} +\frac{l^2}{f_1(r)}dr^{2}  +\frac{r^2}{l^2}\Big(dx^2 + dy^2 + dz^2 \Big)\\
   & f_1(r)=r^2f(r)\nonumber
  \end{align}
 $r$ is the radial coordinate that put us from bulk to the boundary.\\
From linear response theory we know for calculating the shear viscosity we must perturb the metric as $g_{\mu \nu} \to g_{\mu \nu}+\delta g_{xy}$. As we are looking for zero frequency solution, we take the metric perturbation to be $\delta g_{xy}=\frac{r^2}{\l^2}\phi(r)e^{i\omega t}$ and set $\omega=0$.

\begin{equation}\label{metricGB}
ds^{2} =-\frac{f_1(r)}{\l^2} dt^{2} +\frac{\l^2}{f_1(r)} dr^2 + \frac{r^2}{l^2}\Bigg(dx^2 + dy^2 + dz^2 + 2\phi(r)dxdy\Bigg),
\end{equation}
Then by plugging the perturbed metric in the action and keeping terms up to $\phi^2$ we will have,
\begin{equation}
S_2=\frac{-1}{2} \int d^5x \Big(K_1 \phi'^2 -K_2 \phi^2\Big).
\end{equation}
where
\begin{align}
K_1&=\frac{r^5 f_1(r)}{l^3}=\sqrt{-g}g^{rr}=\frac{r}{3l^3}\Big(6 c_0^3 c_3 m^2(r-r_0)+3 c_0^2 c_2m^2(r^2-r_0^2)   \nonumber\\
&+ c_0c_1 m^2(r^3-r_0^3) +\frac{3}{l^2}(r^4-r_0^4)-6e^{2}\ln (\frac{r}{r_0})\Big),\\
K_2&=\frac{1}{2l^3} (2c_0^2 c_2 m^2 r + c_0 c_1 m^2 r^2).
\end{align}
Then the equation of motion will be as follows,
\begin{equation}\label{EoM2}
(K_1 \phi')' + K_2 \phi=0.
\end{equation}
We are going to solve the equation of motion perturbatively in terms of $m^2$ and $e^2$. At first we consider $m^2=e^2=0$. Then the EoM is as follows,
\begin{equation}
(K_3\phi_0')'=0.
\end{equation}
Where $K_3=K_1(m^2=0)$. The solution is as follows,
\begin{equation}\label{sol}
 \phi_0=C_1\int  \frac{dr}{K_3}+C_2.
\end{equation}
By using near horizon approximation we will have,
\begin{equation}
\phi_0(r)=C_2+C_1 (M\log(r-r_0)+...).
\end{equation}
Where $M$ is a constant (depending on $e$ , $l$ and $r_0$) and ellipses are regular terms. Then applying the boundary conditions (regularity at horizon and $\phi=1$ at the boundary) gives $C_1=0$ and $C_2=1$ which means that $\phi_0(r)=1$ is a constant solution. In this case, according to equation (16) in \cite{Hartnoll:2016tri},
\begin{equation}\label{ValuM=0}
\frac{\eta}{s}= \frac{1}{4\pi}\phi(r_0)^2 = \frac{1}{4\pi}
\end{equation}
 Now consider $m^2$  and $e^2$ to be a small parameters and try to solve Eq.(\ref{EoM2}). By Putting $\phi=\phi_0+m^2\phi_1(r)+e^2 \phi_2(r)$ where $\phi_0=1$ and expanding EoM in terms of powers of $m^2$ and $e^2$, we will find,
\begin{align}
&m^2\Big( c_0 c_1 l^2 r^2 + 2c_0^2 c_2 l^2 r +2 r (r^4 - r_0^4) \phi''_1(r)+2(5 r^4- r_0^4) \phi'_1(r)  \Big)=0\\
&e^2\frac{d}{dr}\Big( r (r^4 - r_0^4) \phi'_2(r)\Big)=0
\end{align}
Then the solutions are as follows,
\begin{align}\label{Phi1}
\phi_1(r)&=C_2 -\frac{1}{24 r_0^4}\Big( 2c_0 c_1 l^2 r_0^3 ArcTan\frac{r}{r_0} +24 C_1 \log r +3c_0^2 c_2 l^2 r_0^2 \log(r_0^4-r^4) +\nonumber\\ &+ c_0c_1l^2 r_0^3 \log(r_0-r) - c_0c_1l^2 r_0^3 \log(r_0+r) -6C_1 \log(r^4-r_0^4) \Big).
\end{align}
\begin{equation}\label{Phi2}
\phi_2(r)=C_3+\frac{C_4}{4r_0^8}\bigg(\log(r^4-r_0^4)-4\log r\bigg)
\end{equation}
$\phi(r)$ is as follows,
\begin{align}\label{Phi}
\phi(r)&=\Phi_0- \frac{m^2}{24 r_0^4}\Big( 2c_0 c_1 l^2 r_0^3 ArcTan\frac{r}{r_0} +24 C_1 \log r +3c_0^2 c_2 l^2 r_0^2 \log(r_0^4-r^4)\nonumber\\ &+ c_0c_1l^2 r_0^3 \log(r_0-r) - c_0c_1l^2 r_0^3 \log(r_0+r) -6C_1 \log(r^4-r_0^4) \Big)\nonumber\\&+\frac{e^2C_4}{4r_0^8}\bigg(\log(r^4-r_0^4)-4\log r\bigg)
\end{align}
where $\Phi_0(r)=\phi_0+e^2 C_3+m^2 C_2$.\\  To be regular at the horizon we should get rid of $\log(r-r_0)$ by choosing $C_1=\frac{1}{6}(r_0^3l^2 c_1 c_0 + 3c_0^2 c_2l^2 r_0^2)-\frac{e^2}{m^2r_0^4}C_4$.\\
By substituting $C_4$ in the Eq.(\ref{Phi}) we have,
\begin{align}
\phi(r)&=\Phi_0- \frac{m^2}{24 r_0^4}\Big( 2c_0 c_1 l^2 r_0^3 ArcTan\frac{r}{r_0} +3c_0^2 c_2 l^2 r_0^2 \log(r_0^4-r^4)\nonumber\\ &+ c_0c_1l^2 r_0^3 \log(r_0-r) - c_0c_1l^2 r_0^3 \log(r_0+r) -c_0 c_1 l^2 r_0^3 \log(r^4-r_0^4)\nonumber\\&-3c_0^2 c_2 l^2 r_0^2 \log(r^4-r_0^4)+4c_0 c_1 l^2 r_0^3 \log r+12c^2_0 c_2 l^2 r_0^2 \log r\Big)
\end{align}
The second boundary condition is at the boundary, $\phi(r=\infty)=1$, which gives,
\begin{equation}
\Phi_0= 1+\frac{m^2 l^2 \pi c_0}{24 r_0^2} ( c_1 r_0 + i(3 c_0 c_2 + c_1 r_0))
\end{equation}
Thus we have found $\phi(r)$ everywhere, from horizon to boundary. The solution is as follows,
\begin{align}
\phi(r)&=1+ \frac{m^2 c_0 l^2}{24 r_0^2}\Big((6c_0 c_2 +c_1 r_0)\log(r^2+r_0^2) -2c_1 r_0 ArcTan\frac{r}{r_0}+ \nonumber\\ &+ 2c_1 r_0 \log(r+r_0)-4(3c_0 c_2+c_1 r_0) \log r + \pi c_1 r_0\Big)
\end{align}
Now find it at the horizon, $\phi(r_0)=1+m^2 \phi_1(r_0)+e^2 \phi_2(r_0)$, by taking the limit $r\rightarrow r_0$ and putting the result in viscosity formula we will have,
\begin{align}\label{etas}
 \frac{\eta}{s}&= \frac{1}{4\pi}\phi(r_0)^2 = \frac{1}{4\pi} \big(\Phi_0+m^2 \phi_1(r_0)+e^2 \phi_2(r_0)\big)^2  \nonumber\\
  &= \frac{1}{4\pi} \Bigg(1+\frac{m^2 c_0 l^2}{24 r_0^2}\Big(12 c_0 c_2 \log 2 + c_1 r_0 (\pi +6\log 2)\Bigg).
\end{align}
As mentioned before, we applied these 2 conditions: \\ (i) $\phi$ is regular at horizon $r=r_0$,\\ (ii) goes like to $\phi=1$ near the boundary as $r \to \infty$.\\

By applying the Petrov-like boundary condition the KSS bound will be preserved. The procedure is the same as the second example of \cite{Pan:2016ztm}.

 \section{Conclusion}
\noindent In this paper, applying different boundary conditions gives us different values of $\frac{\eta }{s}$, and subsequently different couplings or theories on the boundary. The Petrov-like boundary \cite{Pan:2016ztm} condition on the hypersurface preserves KSS bound but the Dirichlet boundary and regularity on the horizon conditions violate KSS bound \cite{Hartnoll:2016tri}.
We showed that our result (\ref{etas}) is in agreement to the literature and it is valid perturbatively in $m$ and $e$. When the KSS bound violates it means the model behaves effectively like higher derivative gravity theories \cite{Ref22,Ref23,Ref24,Sadeghi:2015vaa,Parvizi:2017boc} and when the KSS bound saturates it means the model behaves effectively like Einstein-Hilbert gravity.\\

\vspace{1cm}
\noindent {\large {\bf Acknowledgment} }  Author would like to thank Shahrokh Parvizi for useful discussions and Ahmad Moradpur for useful comment and the referee of EPJC for valuable comments which helped to improve the manuscript.



\begin{thebibliography}{}

\bibitem{deRham:2010kj}
C.~de Rham, G.~Gabadadze and A.~J.~Tolley,
Phys.\ Rev.\ Lett.\  {\bf 106}, 231101 (2011)
[arXiv:1011.1232 [hep-th]].

\bibitem{Hassan:2011ea}
  S.~F.~Hassan and R.~A.~Rosen,
  JHEP {\bf 1204}, 123 (2012)
  [arXiv:1111.2070 [hep-th]].


\bibitem{Ref03}
     C.~Brans and R.~H.~Dicke,
      ``Mach's principle and a relativistic theory of gravitation,''  Phys.\ Rev.\  {\bf 124}, 925 (1961).

\bibitem{Ref04}
      M.~Akbar and R.~G.~Cai,
       ``Thermodynamic Behavior of Field Equations for f(R) Gravity,''  Phys.\ Lett.\ B {\bf 648}, 243 (2007)  [gr-qc/0612089].

\bibitem{Ref01}
   D.~Lovelock,
     ``The Einstein tensor and its generalizations,''  J.\ Math.\ Phys.\  {\bf 12}, 498 (1971).


\bibitem{Ref05}
      G.~Cognola, E.~Elizalde, S.~Nojiri, S.~D.~Odintsov, L.~Sebastiani and S.~Zerbini,
        ``A Class of viable modified f(R) gravities describing inflation and the onset of accelerated expansion,''  Phys.\ Rev.\ D {\bf 77}, 046009 (2008)  [arXiv:0712.4017 [hep-th]].

\bibitem{Ref06}
      S.~H.~Hendi,
        ``The Relation between F(R) gravity and Einstein-conformally invariant Maxwell source,''  Phys.\ Lett.\ B {\bf 690}, 220 (2010)  [arXiv:0907.2520 [gr-qc]].

\bibitem{Abbott:2016blz}
  B.~P.~Abbott {\it et al.} [LIGO Scientific and Virgo Collaborations],
  Phys.\ Rev.\ Lett.\  {\bf 116}, no. 6, 061102 (2016)
  [arXiv:1602.03837 [gr-qc]].


\bibitem{Fierz:1939ix}
M.~Fierz and W.~Pauli,
``On relativistic wave equations for particles of arbitrary spin in an electromagnetic field,''
Proc.\ Roy.\ Soc.\ Lond.\ A {\bf 173}, 211 (1939).






\bibitem{Ref1}
   J. M. Maldacena, ``The Large N limit of superconformal field theories and supergravity,'' Int.\ J.\ Theor.\ Phys.\  {\bf 38} (1999) 1113 [Adv.\ Theor.\ Math.\ Phys.\  {\bf 2} (1998) 231] [hep-th/9711200].

\bibitem{Ref2}
 O. Aharony, S.~S.~Gubser, J.~M.~Maldacena, H.~Ooguri and Y.~Oz,
 ``Large N field theories, string theory and gravity,''
  Phys.\ Rept.\  {\bf 323}, 183 (2000)
  [hep-th/9905111].

\bibitem{Ref3}
 J.~Casalderrey-Solana, H.~Liu, D.~Mateos, K.~Rajagopal and U.~A.~Wiedemann,
      ``Gauge/String Duality, Hot QCD and Heavy Ion Collisions,''  arXiv:1101.0618 [hep-th].

\bibitem{Ref4}
D.~Mateos,
  ``String Theory and Quantum Chromodynamics,''
  Class.\ Quant.\ Grav.\  {\bf 24}, S713 (2007)
  [arXiv:0709.1523 [hep-th]].

\bibitem{Ref5}
S.~Bhattacharyya, V.~E.~Hubeny, S.~Minwalla and M.~Rangamani,
 ``Nonlinear Fluid Dynamics from Gravity,''
  JHEP {\bf 0802}, 045 (2008)
  [arXiv:0712.2456 [hep-th]].

\bibitem{Ref6}
  N.~Ambrosetti, J.~Charbonneau and S.~Weinfurtner,
   ``The Fluid/gravity correspondence: Lectures notes from the 2008 Summer School on Particles, Fields, and Strings,''
   arXiv:0810.2631 [gr-qc].

\bibitem{Ref7}
M.~Rangamani,``Gravity and Hydrodynamics: Lectures on the fluid-gravity correspondence,''
  Class.\ Quant.\ Grav.\  {\bf 26}, 224003 (2009)
  [arXiv:0905.4352 [hep-th]].


\bibitem{Ref9}
 J.~Bhattacharya, S.~Bhattacharyya, S.~Minwalla and A.~Yarom,
   ``A Theory of first order dissipative superfluid dynamics,''
   JHEP {\bf 1405}, 147 (2014)
   [arXiv:1105.3733 [hep-th]].

\bibitem{Ref10}
 P. Kovtun,``Lectures on hydrodynamic fluctuations in relativistic theories,''J.\ Phys.\ A {\bf 45} (2012) 473001[arXiv:1205.5040 [hep-th]].



\bibitem{Ref13}
 D.~T.~Son and A.~O.~Starinets,
  ``Viscosity, Black Holes, and Quantum Field Theory,''
  Ann.\ Rev.\ Nucl.\ Part.\ Sci.\  {\bf 57}, 95 (2007)
  [arXiv:0704.0240 [hep-th]].

\bibitem{Ref16}
 G.~Policastro, D.~T.~Son and A.~O.~Starinets,
   ``The Shear viscosity of strongly coupled $ \mathcal{N}=4 $ supersymmetric Yang-Mills plasma,''
   Phys.\ Rev.\ Lett.\  {\bf 87}, 081601 (2001)
   [hep-th/0104066].

\bibitem{Ref17}
 P.~Kovtun, D.~T.~Son and A.~O.~Starinets,
    ``Viscosity in strongly interacting quantum field theories from black hole physics,''
    Phys.\ Rev.\ Lett.\  {\bf 94}, 111601 (2005)
    [hep-th/0405231].

\bibitem{Ref18}
G.~Policastro, D.~T.~Son and A.~O.~Starinets,``From AdS/CFT correspondence to hydrodynamics,''
  JHEP {\bf 0209}, 043 (2002)
  [hep-th/0205052].

\bibitem{Ref19}
 P.~Kovtun, D.~T.~Son and A.~O.~Starinets,
    ``Holography and hydrodynamics: Diffusion on stretched horizons,''
    JHEP {\bf 0310}, 064 (2003)
    [hep-th/0309213].

\bibitem{Shen:2015msa}
  C.~Shen and U.~Heinz,
  Nucl.\ Phys.\ News {\bf 25}, no. 2, 6 (2015)
  doi:10.1080/10619127.2015.1006502
  [arXiv:1507.01558 [nucl-th]]


\bibitem{Sadeghi:2014uqf}
  M.~Sadeghi and S.~Parvizi,
  ``Shear Viscosity to Entropy Density for a Black Brane in 5-dimensional Einstein-Yang-Mills Gravity,''
  arXiv:1411.2358 [hep-th].

\bibitem{Cai:2014znn}
  R.~G.~Cai, Y.~P.~Hu, Q.~Y.~Pan and Y.~L.~Zhang,
``Thermodynamics of Black Holes in Massive Gravity,''
  Phys.\ Rev.\ D {\bf 91}, no. 2, 024032 (2015)
  [arXiv:1409.2369 [hep-th]].

\bibitem{Ref22}
M.~Brigante, H.~Liu, R.~C.~Myers, S.~Shenker and S.~Yaida,
  ``Viscosity Bound Violation in Higher Derivative Gravity,''
  Phys.\ Rev.\ D {\bf 77}, 126006 (2008)
  [arXiv:0712.0805 [hep-th]].


\bibitem{Hartnoll:2016tri}
  S.~A.~Hartnoll, D.~M.~Ramirez and J.~E.~Santos,
  ``Entropy production, viscosity bounds and bumpy black holes,''
  JHEP {\bf 1603}, 170 (2016)
  doi:10.1007/JHEP03(2016)170
  [arXiv:1601.02757 [hep-th]].

\bibitem{Pan:2016ztm}
  W.~J.~Pan and Y.~C.~Huang,
  ``Fluid/gravity correspondence for massive gravity,''
  Phys.\ Rev.\ D {\bf 94}, no. 10, 104029 (2016)
  [arXiv:1605.02481 [hep-th]].

\bibitem{Ref23}
M.~Brigante, H.~Liu, R.~C.~Myers, S.~Shenker and S.~Yaida,
   ``The Viscosity Bound and Causality Violation,''
   Phys.\ Rev.\ Lett.\  {\bf 100}, 191601 (2008)
   [arXiv:0802.3318 [hep-th]].

 \bibitem{Ref24}
  I.~P.~Neupane and N.~Dadhich,
     ``Entropy Bound and Causality Violation in Higher Curvature Gravity,''
     Class.\ Quant.\ Grav.\  {\bf 26}, 015013 (2009)
     [arXiv:0808.1919 [hep-th]].


\bibitem{Sadeghi:2015vaa}
  M.~Sadeghi and S.~Parvizi,
  ``Hydrodynamics of a black brane in Gauss-Bonnet massive gravity,''
  Class.\ Quant.\ Grav.\  {\bf 33}, no. 3, 035005 (2016)
  [arXiv:1507.07183 [hep-th]].


\bibitem{Parvizi:2017boc}
  S.~Parvizi and M.~Sadeghi,
  ``Holographic Aspects of a Higher Curvature Massive Gravity,''
  arXiv:1704.00441 [hep-th].
\end{thebibliography}
\end{document}